\documentclass[a4paper, aps,prl,showpacs,floatfix
,twocolumn,superscriptaddress]{revtex4}

\usepackage{graphicx}  
\usepackage{amsmath, amssymb}

\newcommand{\ket}[1]{\vert#1\rangle} 
\newcommand{\bra}[1]{\langle#1\vert}

\newcommand{\proj}[1]{\ket{#1}\bra{#1}}
\newcommand{\mean}[1]{\langle #1 \rangle}
\newcommand{\abs}[1]{\lvert#1\rvert}
\newcommand{\Abs}[1]{\big\lvert#1\big\rvert}
\newcommand{\norm}[1]{\lVert#1\rVert}
\newcommand{\Norm}[1]{\big\lVert#1\big\rVert}
\DeclareMathOperator{\tr}{Tr}
\newcommand{\identity}{\mathbb{I}}
\renewcommand{\S}{\mathbb{S}}

\newcommand{\R}{\mathbb{R}}
\newcommand{\Hil}{{\cal H}}

\usepackage{color}

\begin{document}

\title{Universality in open system entanglement dynamics}

\author{Markus Tiersch}
\affiliation{%
Physikalisches Institut der Albert--Ludwigs--Universit\"at,
Freiburg, Germany
}
\affiliation{%
Institute for Quantum Optics and Quantum Information,
Innsbruck, Austria}
\author{Fernando de Melo}
\affiliation{%
Physikalisches Institut der Albert--Ludwigs--Universit\"at,
Freiburg, Germany
}
\affiliation{%
Instituut voor Theoretische Fysica, Katholieke Universiteit Leuven,
Belgium
}
\author{Andreas Buchleitner}
\affiliation{%
Physikalisches Institut der Albert--Ludwigs--Universit\"at,
Freiburg, Germany
}




\begin{abstract}
We show that the entanglement evolution of an open quantum system is the same for the vast majority of initial pure states, in the limit of large Hilbert space dimensions.
\end{abstract}

\maketitle


The ability to characterize and control quantum systems with an ever increasing number of constituents is an ever more desired and necessary skill in many fields of modern physics~\cite{ref:modern}, materials science~\cite{ref:material}, and even biology~\cite{ref:Qbio}.
From a strictly deterministic point of view, however, the rapidly growing generic complexity of such large systems apparently renders this task ineffable. Yet, similarly to thermodynamics, a \emph{statistical} description often allows for the extraction of robust, generic features which emerge in the limit of large system size, and imply quantitative predictions.

Entanglement, an unmistakeable quantum signature, is a prime example of the above, apparent illusiveness of an exhaustive characterization as the system size is scaled up:
The structure of many-particle entanglement turns more and more intricate with the exponentially increasing number of possible correlations between different subgroups of particles. Thus, a complete characterization of a large, composite quantum state requires an experimental overhead that increases exponentially with the number of system constituents.
Even worse, entanglement tends to get ever more fragile when enlarging the system size: the more degrees of freedom, the more difficult it becomes to shield quantum coherences, which are necessary for entanglement, against the detrimental influence of a noisy environment.
In such situations, the strong quantum correlations due to entanglement additionally need to be distinguished from classical correlations induced by the ambient noise. 
This is in general accomplished by high dimensional optimization procedures on the space of all quantum states~\cite{ref:Uhlmann,ref:distance}, leaving little hope for quantitative predictions on entanglement evolution in large, and noisy systems.
On the other hand, the signatures of entanglement in such systems are of high fundamental and, potentially, practical interest, e.g.\ when it comes to harnessing the computational power of quantum algorithms~\cite{ref:onewayQC} or assessing the role of quantum correlations in intrinsically noisy biological systems~\cite{ref:Qbio}. It is therefore a key issue and the subject of the present Letter to estimate the characteristic time scales in which entanglement is present in such adverse situations.

Here we consider the fate of entangled states of large quantum systems, which require a high dimensional Hilbert space, in contact with an incoherent environment. We show that a statistical analysis over \emph{generic} initial states unveils \emph{universal} -- state independent -- open system entanglement evolution in the limit of large Hilbert space dimensions. In this thermodynamic limit, an efficient characterization of entanglement dynamics is thus again possible.

\begin{figure}
	\centering
	\includegraphics[width=0.8\linewidth]{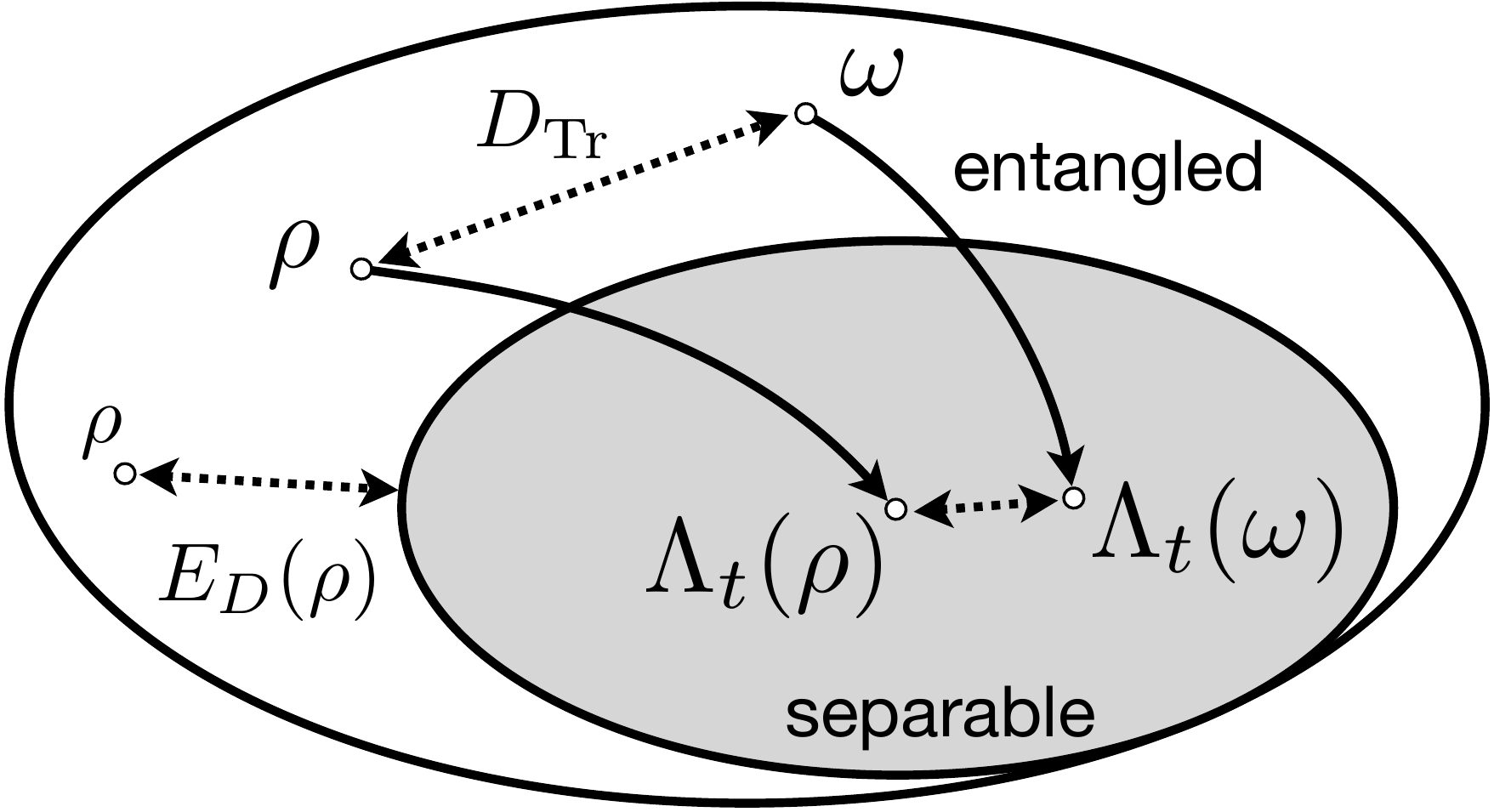}
	\caption{State space is a convex set containing all quantum states of a given dimension. Open system dynamics due to a map $\Lambda_t$ cannot increase the distance between any two states (as measured by the trace distance). Entanglement measures, as $E_D$, monotonically increase with the distance to the set of separable states (in gray).}
	\label{fig:statespace}
\end{figure}

%
More specifically, let us start with a composite quantum system in a pure state, characterized by a vector $\ket{\chi}$ in a Hilbert space $\Hil = \Hil_A\otimes\Hil_B\dotsm\otimes\Hil_N$ with dimension $d = d_A d_B\dotsm d_N$.
The system then undergoes some dynamics by virtue of its Hamiltonian, and also couples to uncontrollable degrees of freedom, which define its environment (denoted by subscript $E$).
When focusing on the system alone, its state $\rho$ evolves as  described by a time-dependent map $\Lambda_t$:
\begin{equation}
\rho(0) \mapsto \rho(t)=\tr_E \big[ U(t) \rho_{SE}(0) U(t)^\dag \big] =: \Lambda_t [\rho(0)],
\end{equation}
where $U(t)$ is the unitary time evolution operator obtained as solution of the Schr\"odinger equation for system \emph{and} environment (subscript $SE$).
Such maps $\Lambda_t$ are only independent of the system's initial state $\rho(0)$ if system and environment are initially uncorrelated or at most classically correlated~\cite{Shabani:2009}.
This is the case for many implementations of quantum information tasks~\cite{ref:purestates}, where initial system state are almost pure, and therefore uncorrelated with the environment in very good approximation.
In what follows, we thus assume the initial state of system and environment to be of the form $\rho_{SE}(0)=\proj{\chi}\otimes\rho_E(0)$.
Since, beyond this latter factorization, our ansatz makes no assumption on the specific form of the Hamiltonian and thus of $U(t)$, and neither of $\rho_E(0)$, our subsequent results are applicable for arbitrary open system dynamics of the given initial states, including non-Markovian effects that may arise in the course of the evolution.
The following derivations are solely grounded in few geometrical properties of the set of states, and properties of entanglement measures thereon. The next steps and properties needed to prepare our main result are visualized in Fig.~\ref{fig:statespace}.

As a measure for the relative effect of the open system dynamics $\Lambda_t$ onto two states $\rho$ and $\omega$, we choose the metric distance
\begin{equation}
D_{\tr}(\rho,\omega) = \norm{\rho-\omega}_{\tr}
:= \tr \abs{ \rho-\omega },
\end{equation}
induced by the trace norm $\norm{\cdot}_{\tr}$ on state space.
The distance of two states measured by $D_{\tr}$ decreases monotonically upon application of any such $\Lambda_t$~\cite{ref:RuskaiTraceDist},
\begin{equation} \label{eq:ChannelContraction}
D_{\tr}[\Lambda_t(\rho),\Lambda_t(\omega)] \leq \eta_{\Lambda_t} \, D_{\tr}(\rho,\omega)\,
\quad \text{with} \quad
\eta_{\Lambda_t} \leq 1.
\end{equation}
Note that, for \emph{closed} dynamics, i.e., without coupling to the environment, equality holds with $\eta_{\Lambda_t}=1$, as a consequence of the unitarity of quantum mechanics.
\emph{Open} dynamics, however, introduces additional mixing of the system state, such that the state space is effectively contracted.
This dynamical effect is captured by the change of $\eta_{\Lambda_t}$.
For example, when coupled to a thermal bath, all initial states converge to thermal equilibrium, and $\eta_{\Lambda_t}$ approaches zero asymptotically in time.

Turning to entanglement, we do not focus on a specific entanglement quantifier, but rather only require that it slowly varies on the set of states.
More specifically, we demand a ``strong'' form of continuity, namely Lipschitz continuity~\cite{ref:Rockafellar},
\begin{equation} \label{eq:LipschitzMeasure}
\Abs{E(\rho)-E(\omega)} \leq \eta_E \, D_{\tr}(\rho,\omega)\, ,
\end{equation}
where $\eta_E$ is called the Lipschitz constant. 
Examples of such entanglement measures $E(\rho)$ are the ones defined as the minimum distance of the state $\rho$ to the set $\mathcal{S}$ of separable states: $E_D(\rho) = \min_{\sigma \in \mathcal{S}} D(\rho,\sigma)$~\cite{ref:distance}.
All distances $D$ which abide to the dynamic monotonicity condition~\eqref{eq:ChannelContraction}~\cite{ref:FootnoteDistance}, fulfill the Lipschitz requirement with constant $\eta_{E_D}=1$, by virtue of the triangle inequality.
Another example is negativity $\mathcal{N}$, a computable entanglement monotone~\cite{ref:Negativity}, but with a different constant $\eta_\mathcal{N}$ (see the appendix).

Under conditions~(\ref{eq:ChannelContraction}) and~(\ref{eq:LipschitzMeasure}), we can asses how much two initially pure states differ in their entanglement after exposure to the same incoherent dynamics.
For states $\rho(t)=\Lambda_t(\proj{\chi})$ and $\omega(t)=\Lambda_t(\proj{\psi})$ we obtain
\begin{align} \label{eq:differenceEstimate}
\Abs{E(\rho)-E(\omega)} &\leq
\eta_E \, D_{\tr}\big[\Lambda_t(\proj{\chi}),\Lambda_t(\proj{\psi})\big] \nonumber \\
 &\leq \eta_E\, \eta_{\Lambda_t} \, D_{\tr}\big[\proj{\chi},\proj{\psi}\big] \nonumber \\
 &\leq 2\eta_E\, \eta_{\Lambda_t} \, \Norm{\ket{\chi}-\ket{\psi}}. 
\end{align}
That is, the entanglement quantifier $E$ inherits Lipschitz continuity, with constant $2\eta_E\,\eta_{\Lambda_t}$, even with respect to the Euclidean distance between the two initial state vectors in $\mathcal{H}$.
The difference in entanglement of the two mixed final states is essentially bounded by the distance between the initial states.
With the knowledge of the entanglement of a \emph{single} probe state, say of $\rho(t)$, for an initial state $\ket{\chi}$, we are able to predict the entanglement evolution of any other pure initial state $\ket{\psi}$ within an error margin given by \eqref{eq:differenceEstimate} -- this without the need to evolve the state, and to compute the resulting mixed state's entanglement.

Rather than studying specific instances of initial states, we aim at a statement about \emph{generic} pure states in a statistical manner.
To avoid any bias, we use random states which uniformly cover the space of pure states.
States that exhibit the properties of random states naturally appear in registers of quantum computers after long sequences of gates~\cite{gates}, in quantum systems with a chaotic classical counterpart~\cite{chaos}, and also subsets of pure states, e.g.\ graph states which emerge in spin gases~\cite{spinGas}, show the properties of a uniform distribution~\cite{Plato}.
In addition, for many interesting scenarios of biology and chemistry such as transport phenomena in proteins, the initial state and/or the Hamiltonian are not exactly known, e.g.~\cite{ref:FMOdynamics}. In these cases a uniform distribution of initial states in the potential subspace is the best prior.

%
Let us first note that the set of state vectors in $\mathcal H$ is isomorphic to a unit sphere $\S^{2d-2}$ in $\R^{2d-1}$, where the real and imaginary part of the expansion coefficients into any basis constitute the coordinates, constrained by normalization and ignoring the global phase by choosing the first component real.
With the above preparation, we can now infer our central result by employing the well-known Levi's lemma~\cite{ref:Ledoux,ref:Milman,ref:Hayden}:
The probability for a deviation larger than $\epsilon$ of $E[\rho(t)] := E[\Lambda_t (\ket{\chi}\bra{\chi})]$ from its mean $\mean{E}(t) := \int d\psi\ E[\Lambda_t(\ket{\psi}\bra{\psi})]$ over all initial states, given the uniform initial distribution on $\S^{2d-2}$, exhibits an exponential suppression:
(i) in the deviation $\epsilon$
, and moreover, 
(ii) in the system dimension $d$:
\begin{equation} \label{eq:Concentration}
\Pr \big( \Abs{E[\rho(t)] - \mean{E}(t)} > \epsilon \big)
\leq
4\exp \left( -C \frac{2d-1}{4\eta_E^2 \eta_{\Lambda_t}^2} \epsilon^2 \right).
\end{equation}
The constant $C$ can be chosen $(24 \,\pi^2)^{-1}$~\cite{diss_tiersch}. This holds for all entanglement quantifiers $E$ that fulfill~\eqref{eq:LipschitzMeasure}, even when quantifying multipartite entanglement.

%
\begin{figure*}
	\centering
	\includegraphics[width=0.9\textwidth]{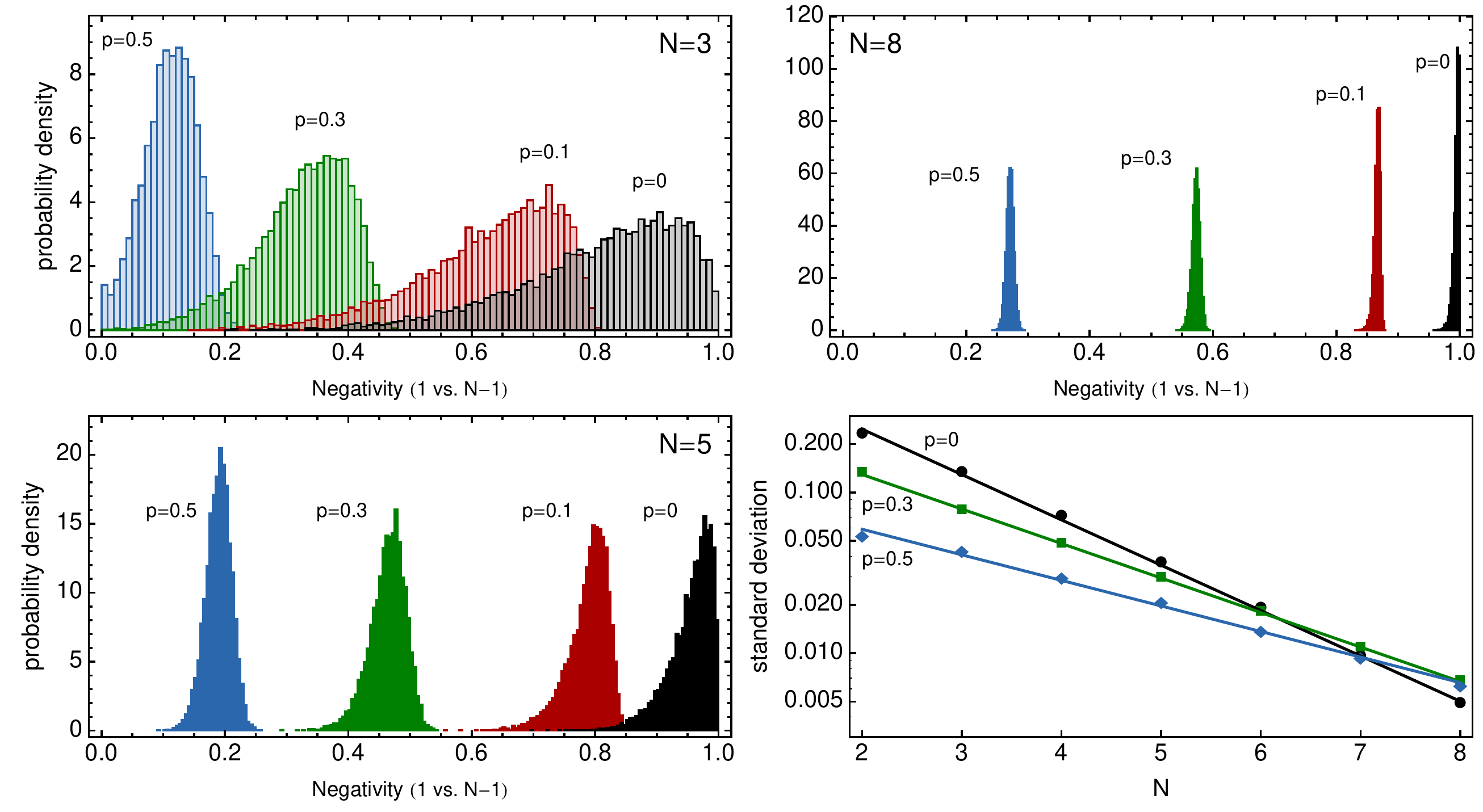}
	\caption{(Color online) Distributions of negativity.
	The data for negativity is obtained for the least balanced partition (one vs.\ $N-1$ qubits)
	of $N=3,5,8$ 
	qubits, at different snapshots of the evolution.
	We sample over 10\,000 uniformly distributed pure initial states of $N$ qubits
	which evolve under local coupling to a decoherence reservoir.
	The system--environment interaction time $t$ is parametrized
	by $p = (1 - e^{-\Gamma t})$, with $\Gamma$ the local decoherence rate.
	Clearly, the bigger $N$, the more concentrates the negativity distribution
	around its mean value, for all times, as exemplified in the logarithmically plotted standard deviation for system sizes $N=2,\dotsc,8$ and three time steps of the evolution $p=0, 0.3, 0.5$ (bottom right). Lines are linear fits to the data.}
	\label{fig:distributions}
\end{figure*}
%

As a concrete application of this result, we consider the normalized negativity $\mathcal{N}/\mathcal{N_\text{max}}$ of a bipartite system on $\Hil_A\otimes \Hil_B$. Its Lipschitz constant (see appendix) reads $\eta_{(\mathcal{N}/\mathcal{N}_\text{max})} = d_A/(d_A-1)$, and the following inequality for a deviation from the mean entanglement holds:
\begin{widetext}
\begin{equation} \label{eq:ConcentrationNeg}
\Pr \left( \left\vert \frac{\mathcal{N}\big[\rho(t)\big] - \mean{\mathcal{N}}(t) }{ \mathcal{N}_\text{max}} \right\vert > \epsilon \right)
\leq
4 \exp \left( -C \frac{(2 d_A d_B - 1)(d_A - 1)^2}{4 d_A^2 \, \eta_{\Lambda_t}^2} \epsilon^2 \right).
\end{equation}
\end{widetext}
This expresses the concentration effect visualized in Fig.~\ref{fig:distributions} for ensembles of $N=2,\dots,8$ qubits (two-level systems), each locally coupled to an environment that destroys only the coherences.
Negativity is here evaluated with respect to the least balanced bipartition of the qubit register, i.e., it quantifies the entanglement of one qubit ($d_A=2$) with the remaining $N-1$ qubits ($d_B=2^{N-1}$), after application of the decoherence dynamics.
We uniformly sample 10\,000 pure initial states, and parametrize the dynamics by the probability $p$ for complete decoherence of a single qubit.
Within the Markov approximation $p = 1 - e^{-\Gamma t}$, where $\Gamma$ is the decoherence rate~\cite{ref:FootnotePhaseDamping}.
The $\log$-plot of the variances for increasing system size, shown at the bottom right of Fig.~\ref{fig:distributions}, establishes the exponential concentration around the mean.
Thereby, our numerical results underpin the \emph{mixed state entanglement concentration} predicted by \eqref{eq:Concentration} and \eqref{eq:ConcentrationNeg}.
Furthermore, the concentration observed by the numerical experiment is even stronger than estimated -- the measured variances are approximately one order of magnitude smaller than the variance which we would infer from (\ref{eq:ConcentrationNeg}), with our above estimate of $C$ and the Lipschitz constants.

The above results determine the open system evolution of entanglement with an error \emph{exponentially small} in the dimension of the underlying Hilbert space (provided $\eta_E$ does not increase faster than $\sqrt{d}$).
Therefore, in high dimensions, it suffices to monitor the entanglement evolution of a \emph{single}, generic pure state, in order to predict the fate of any other typical state subject to the same dynamics.
Qualitatively different entanglement dynamics for different initial states~\cite{ref:ESD} will occur only as singular effects in sufficiently large quantum systems.

Inasmuch as knowledge of a single state's entanglement evolution fully determines (in the present, asymptotic sense) the result for arbitrary states, our result is reminiscent of previously derived entanglement evolution equations~\cite{ref:Fact,ref:FactDxD}, where the final entanglement of an arbitrary initial pure state was shown to be fully characterized by the entanglement evolution of a maximally entangled state.

The \emph{concentration} of open system entanglement evolution, as spelled out by \eqref{eq:Concentration} and \eqref{eq:ConcentrationNeg}, has yet another bearing for the optimization problem routinely encountered~\cite{ref:Uhlmann,ref:distance} when evaluating the entanglement of arbitrary mixed states.
For mixed states generated through arbitrary \emph{physical} dynamics $\Lambda_t$ from a uniform distribution of pure states, mixed state entanglement concentration suggests a reduction of the optimization space:
A single representative of the obtained sample \emph{generated under a specific physical evolution}, selected with convenient properties (symmetries in terms of its pure state decomposition), will suffice to impose exponentially narrow constraints on the optimization for all other states.

{\it Acknowledgements.} We enjoyed discussions with Mark Fannes, Alexander Holevo, and particularly thank Reinhard Werner for discussions on the Lipschitz constant of negativity.
We also profited from exchanges with Carlos Viviescas, Alejo Sales, and Daniel Cavalcanti, and express our gratitude for the hospitality of Berge Englert and the NUS, where part of the work was done, as well as Luiz Davidovich's group at the Universidade Federal do Rio de Janeiro, where this work was finished.
DAAD/CAPES is acknowledged for financial support through the PROBRAL program.
M.T. was partially supported by the IMS.
F.M. acknowledges support by the Alexander von Humboldt Foundation, and Belgium Interuniversity Attraction Poles Programme P6/02.



{\em Appendix.} In order to calculate the Lipschitz constant for the negativity of a bipartite system, we use its definition through the trace norm of the partially transposed state~\cite{ref:Negativity},
$\mathcal{N}(\rho) := ( \norm{(\identity\otimes T)(\rho)}_{\tr} - 1 )/ 2$, 
where $T$ represents the transposition acting just on $\Hil_B$.
By means of the triangle inequality, $\Abs{\norm{x}-\norm{y}}\leq\norm{x-y}$, and the linearity of the partial transpose we arrive at a first estimation for states $\rho$ and $\omega$:
\begin{equation}
\abs{\mathcal{N}(\rho)-\mathcal{N}(\omega)}
\leq
\frac{1}{2} \Norm{ (\identity\otimes T)(\rho-\omega)}_{\tr}
\,.
\end{equation}
The remaining norm can be estimated with the operator norm defined as 
$\norm{A}_\text{op} := \sup_x (\norm{Ax} / \norm{x})$.
By choosing a particular $x$, not necessarily optimal, we obtain a lower bound:
$
\norm{(\identity\otimes T)}_\text{op}
\geq
\norm{(\identity\otimes T)(\rho-\omega)}_{\tr} / \norm{\rho-\omega}_{\tr}\,.
$
This leads to the next estimate for negativity:
\begin{equation}
\abs{\mathcal{N}(\rho)-\mathcal{N}(\omega)}
\leq
\frac{1}{2} \norm{(\identity\otimes T)}_\text{op} \norm{\rho-\omega}_{\tr}
\,.
\end{equation}
The maximization of the partial transposition is obtained with a maximally entangled state, which yields $\norm{(\identity\otimes T)}_\text{op}=d_A$, with $d_A$ the dimension of the smallest subsystem of the bipartition.
Consequently, we find Lipschitz continuity for negativity,
\begin{equation}
\abs{\mathcal{N}(\rho)-\mathcal{N}(\omega)}
\leq
\frac{d_A}{2} \norm{\rho-\omega}_{\tr}
\,,
\end{equation}
with Lipschitz constant $\eta_\mathcal{N}=d_A/2$.
For the normalized negativity, i.e., rescaled with respect to its maximal value $\mathcal{N}_\text{max}=(d_A-1)/2$, we finally obtain $\eta_{(\mathcal{N}/\mathcal{N}_\text{max})}= d_A/(d_A-1)$.



\end{document}